\documentclass[12pt]{article}
\usepackage{amsmath,amssymb,amsthm,amsxtra,overpic,bbm,bm,epsfig,subfigure}
\usepackage{hyperref}
\usepackage{mathrsfs}
\usepackage{enumitem}
\usepackage{graphicx}
\usepackage{color}
\usepackage{comment}
\usepackage{epstopdf}
\usepackage{float}
\usepackage{cite}
\usepackage{slashed,stmaryrd}

\def\thefootnote{\fnsymbol{footnote}}

\usepackage{hyperref}
\usepackage{slashed,stmaryrd}

\usepackage{lscape}%
\usepackage{array}
\usepackage{booktabs}%

\begin{document}

\vspace{0.2cm}

\begin{center}
{\Large\bf Neutrino Masses, Leptonic Flavor Mixing and Muon $(g-2)$ in the Seesaw Model with the ${\rm U}(1)^{}_{L^{}_\mu - L^{}_\tau}$ Gauge Symmetry}
\end{center}

\vspace{0.2cm}

\begin{center}
{\bf Shun Zhou}~\footnote{E-mail: zhoush@ihep.ac.cn}
\\
\vspace{0.2cm}
{\small
Institute of High Energy Physics, Chinese Academy of Sciences, Beijing 100049, China\\
School of Physical Sciences, University of Chinese Academy of Sciences, Beijing 100049, China\\}
\end{center}

\vspace{1.5cm}

\begin{abstract}
The latest measurements of the anomalous muon magnetic moment $a^{}_\mu \equiv (g^{}_\mu - 2)/2$ show a $4.2\sigma$ discrepancy between the theoretical prediction of the Standard Model and the experimental observations. In order to account for such a discrepancy, we consider a possible extension of the type-(I+II) seesaw model for neutrino mass generation with a gauged $L^{}_\mu - L^{}_\tau$ symmetry. By explicitly constructing an economical model with only one extra scalar singlet, we demonstrate that the gauge symmetry ${\rm U}(1)^{}_{L^{}_\mu - L^{}_\tau}$ and its spontaneous breaking are crucially important not only for explaining the muon $(g - 2)$ result but also for generating neutrino masses and leptonic flavor mixing. Various phenomenological implications and experimental constraints on the model parameters are also discussed.
\end{abstract}


\def\thefootnote{\arabic{footnote}}
\setcounter{footnote}{0}

\newpage

\section{Introduction}\label{sec:intro}

Based on the quantum field theories with gauge symmetries, the Standard Model (SM) has proved to be extremely successful in the description of strong, electromagnetic and weak interactions among elementary particles in nature. However, the fermion mass spectra, flavor mixing patterns and CP violation have been completely unexplained in the SM~\cite{Xing:2019vks}. In particular, the discovery of neutrino oscillations calls for new physics beyond the SM to accommodate tiny neutrino masses and significant leptonic flavor mixing. Apart form the direct searches for new physics at the high-energy frontiers, another important approach is to precisely measure the basic properties of the SM particles and look for clear deviations from the SM predictions.

Very recently, the Muon $(g-2)$ Collaboration at Fermi National Laboratory in the US has released the precise measurement of the anomalous muon magnetic moment $a^{}_\mu \equiv (g^{}_\mu - 2)/2$ and found a $4.2\sigma$ discrepancy with the SM prediction~\cite{Abi:2021gix}, when combined with the final result from the E821 experiment at Brookhaven National Laboratory in 2006~\cite{Bennett:2006fi}. The reported difference between the combined experimental result $a^{\rm exp}_\mu = 116\ 592\ 061(41) \times 10^{-11}$~\cite{Abi:2021gix} and the SM prediction $a^{\rm SM}_\mu = 116\ 591\ 810(43) \times 10^{-11}$~\cite{Davier:2010, Hagiwara:2011af, Davier:2017zfy, Colangelo:2018mtw, Hoferichter:2019mqg, Davier:2019can, Keshavarzi:2019abf, Kurz:2014wya, FermilabLattice:2017wgj, Budapest-Marseille-Wuppertal:2017okr, RBC:2018dos, Giusti:2019xct, Shintani:2019wai, FermilabLattice:2019ugu, Gerardin:2019rua, Aubin:2019usy, Giusti:2019hkz, Melnikov:2003xd, Masjuan:2017tvw, Colangelo:2017fiz, Hoferichter:2018kwz, Gerardin:2019vio, Bijnens:2019ghy, Colangelo:2019uex, Pauk:2014rta, Danilkin:2016hnh, Knecht:2018sci, Eichmann:2019bqf, Roig:2019reh, Colangelo:2014qya, Blum:2019ugy, Aoyama:2012wk, Aoyama:2019ryr, Czarnecki:2002nt, Gnendiger:2013pva, Aoyama:2020ynm} of anomalous muon magnetic moment is
\begin{eqnarray}
\Delta a^{}_\mu \equiv a^{\rm exp}_\mu - a^{\rm SM}_\mu = 251(59) \times 10^{-11} \; ,
\label{eq:Deltaa}
\end{eqnarray}
where the $1\sigma$ error in the last two digits is given by the number in the parentheses. Although such a discrepancy needs to be further scrutinized by more experimental data and by more reliable calculations~\cite{Borsanyi:2020mff}, it has already stimulated a great number of theoretical works on the new physics interpretations and their connections to other fundamental problems in particle physics, astronomy and cosmology~\cite{Lindner:2016bgg, Hiller:2019mou, Cvetic:2020vkk, Chen:2020tfr, Borah:2020jzi, Arbelaez:2020rbq, Jana:2020joi, Yin:2020afe, Yin:2021yqy, Chiang:2021pma, Athron:2021iuf, Babu:2021jnu, Borah:2021jzu, Arcadi:2021yyr, Chen:2021jok, Nomura:2021oeu, Endo:2021zal, Zhang:2021gun, Yang:2021duj, Aboubrahim:2021rwz, Escribano:2021css, Ahmed:2021htr, Li:2021poy, Ferreira:2021gke, Zu:2021odn, Chun:2021dwx, Chen:2021vzk, Bai:2021bau, Cox:2021gqq, Cao:2021tuh, Calibbi:2021qto, Han:2021ify, Cadeddu:2021dqx, Gu:2021mjd, Buen-Abad:2021fwq, Han:2021gfu, Abdughani:2021pdc, VanBeekveld:2021tgn, Wang:2021bcx, Zhu:2021vlz, Wang:2021fkn, Baum:2021qzx, Brdar:2021pla, Ge:2021cjz, Yin:2021mls, Arcadi:2021cwg}. See, e.g., Refs.~\cite{Lindner:2016bgg, Athron:2021iuf}, for an excellent summary of new-physics scenarios and earlier references.

Motivated by this exciting progress, we propose a simple but viable model to simultaneously account for neutrino masses, leptonic flavor mixing and the anomalous muon magnetic moment. The basic idea is to augment the type-(I+II) seesaw model~\cite{Minkowski, Yanagida, Gell-Mann, Glashow, Mohapatra, Schechter:1980gr, Cheng:1980qt} by introducing the ${\rm U}(1)^{}_{L^{}_\mu - L^{}_\tau}$ gauge symmetry~\cite{Foot:1990mn, He:1991qd, Foot:1994vd} and an additional singlet scalar field $S$. The results obtained in the present work are distinct from those in the previous works at least in two aspects. First, the flavor-dependent gauge symmetry ${\rm U}(1)^{}_{L^{}_\mu - L^{}_\tau}$ places very restrictive constraints on the lepton flavor structures, so it is nontrivial to correctly reproduce in a simple way two neutrino mass-squared differences and three flavor mixing angles as observed in neutrino oscillation experiments~\cite{Heeck:2011wj, Araki:2012ip, Heeck:2012cd, Baek:2015mna, Patra:2016shz, Biswas:2016yan, Asai:2017ryy, Chen:2017gvf, Nomura:2018cle, Asai:2019ciz, Majumdar:2020xws}. In our model, the singlet scalar field $S$ and the triplet Higgs $\Delta$ are assigned with the opposite charges $Q^{}_{\mu\tau}(S) = +1$ and $Q^{}_{\mu\tau}(\Delta) = - 1$ under the ${\rm U}(1)^{}_{L^{}_\mu - L^{}_\tau}$ symmetry, respectively. As a consequence, the effective Majorana mass matrix $M^{}_\nu$ of three light neutrinos can take the two-zero texture ${\bf B}^{}_3$~\cite{Frampton:2002yf, Xing:2002ap, Xing:2002ta}, which leads to readily testable predictions for neutrino mixing parameters~\cite{Fritzsch:2011qv, Zhou:2015qua}. Second, we point out that both the contributions from the Higgs triplet and those from heavy Majorana neutrinos to $\Delta a^{}_\mu$ appear with a negative sign. Therefore, the observed discrepancy $\Delta a^{}_\mu$ can only be explained by the contribution from the gauge boson $Z^\prime$ associated with the ${\rm U}(1)^{}_{L^{}_\mu - L^{}_\tau}$ gauge symmetry. As there is no tree-level interaction between electrons and $Z^\prime$, the anomalous electron magnetic moment $\Delta a^{}_e \equiv a^{\rm exp}_e - a^{\rm SM}_e$ receives the dominant correction from the Higgs triplet, implying a negative sign of $\Delta a^{}_e$ that remains consistent with experimental observation at the $2\sigma$ level. Our model offers an explicit example to realize the opposite signs for $\Delta a^{}_\mu$ and $\Delta a^{}_e$ in the same framework.

The remaining part of this work is organized as follows. In Sec.~\ref{sec:framework}, we introduce the extended version of type-(I+II) model with a gauged $L^{}_\mu - L^{}_\tau$ symmetry, and explain its main features. The phenomenological implications for neutrino masses and flavor mixing are discussed in Sec.~\ref{sec:mass}, while the anomalous electron and muon magnetic moments are calculated in Sec.~\ref{sec:mag}. Finally, we make some concluding remarks in Sec.~\ref{sec:conclu}.

\section{The Seesaw Model with ${\rm U}(1)^{}_{L^{}_\mu - L^{}_\tau}$ Symmetry} \label{sec:framework}

Although the individual lepton numbers $L^{}_\alpha$ (for $\alpha = e, \mu, \tau$) and the baryon number $B$ happen to be conserved in the SM at the classical level, they are actually violated by quantum anomalies~\cite{tHooft:1976rip}. However, the differences between any two of these global charges, such as $B/3 - L^{}_\alpha$ (for $\alpha = e, \mu, \tau$), $L^{}_e - L^{}_\mu$ and $L^{}_\mu - L^{}_\tau$, are anomaly-free and can naturally be promoted to gauge symmetries. The well-established phenomena of neutrino oscillations, in particular the approximate $\mu$-$\tau$ exchange or reflection symmetries observed in the neutrino flavor mixing~\cite{Xing:2015fdg}, indicate that the ${\rm U}(1)^{}_{L^{}_\mu - L^{}_\tau}$ gauge symmetry can be compatible with experimental observations.

In the canonical type-(I+II) seesaw model, where three right-handed neutrino singlets $\nu^{}_{\alpha {\rm R}}$ (for $\alpha = e, \mu, \tau$) and one Higgs triplet $\Delta$ under the ${\rm SU}(2)^{}_{\rm L}$ gauge symmetry are added into the SM particle content, it is natural to obtain tiny Majorana masses of three ordinary neutrinos. However, the flavor structures of the lepton Yukawa couplings and right-handed neutrino mass matrix are entirely unconstrained in the type-(I+II) seesaw model. As we shall see in the next section, the ${\rm U}(1)^{}_{L^{}_\mu - L^{}_\tau}$ gauge symmetry could impose tight constraints on the lepton flavor structures. Usually the flavor structures are constrained to be too simple to accommodate two neutrino mass-squared differences $\Delta m^2_{ij} \equiv m^2_i - m^2_j$ (for $ij = 21, 31$) and three neutrino mixing angles $\{\theta^{}_{12}, \theta^{}_{13}, \theta^{}_{23}\}$ extracted from neutrino oscillation experiments. For this reason, we have to modify the canonical type-(I+II) seesaw model in the most economical way by bringing in an additional complex scalar $S$, which is singlet under the SM gauge symmetry but possesses a nontrivial ${\rm U}(1)^{}_{L^{}_\mu - L^{}_\tau}$ charge. The relevant fields and their charges under the overall gauge symmetry ${\rm SU}(2)^{}_{\rm L}\otimes {\rm U}(1)^{}_{\rm Y}\otimes {\rm U}(1)^{}_{L^{}_\mu - L^{}_\tau}$ have been summarized in the Table \ref{Table: model}. In the following, we explain the extra terms appearing in the gauge-invariant Lagrangian and briefly discuss their phenomenological implications.

First of all, there is a new gauge symmetry, for which the gauge field will be denoted as $Z^\prime_\mu$ and the gauge boson as $Z^\prime$. In addition to the new kinetic term~\footnote{In general, there could also exist a mixing term $-\epsilon Z^\prime_{\mu\nu} B^{\mu\nu}/4$ between the ${\rm U}(1)^{}_{\rm Y}$ and ${\rm U}(1)^{}_{L^{}_\mu - L^{}_\tau}$ gauge bosons in the Lagrangian. Even if this mixing is assumed to be vanishing initially (i.e., $\epsilon = 0$), it will be generated at the loop level by the particles that are charged under both ${\rm U}(1)^{}_{\rm Y}$ and ${\rm U}(1)^{}_{L^{}_\mu - L^{}_\tau}$. For simplicity, we assume that the mixing between $Z^\prime$ and the SM neutral gauge bosons is radiatively generated and thus highly suppressed.} composed of the field-strength tensor $Z^\prime_{\mu\nu} \equiv \partial^{}_\mu Z^\prime_\nu - \partial^{}_\nu Z^\prime_\mu$, the covariant derivative will be modified to
\begin{eqnarray}
    D^\prime_\mu \equiv \partial^{}_\mu - i g \tau^a W^a_\mu - i g^\prime \frac{Y}{2} B^{}_\mu - i g^{}_{Z^\prime} Q^{}_{\mu\tau} Z^\prime_\mu \; ,
    \label{eq:cov}
\end{eqnarray}
where $Q^{}_{\mu\tau}$ stands for the charge under the ${\rm U}(1)^{}_{L^{}_\mu - L^{}_\tau}$ symmetry (cf. the charges in the last row of Table~\ref{Table: model}), and $g^{}_{Z^\prime}$ for the corresponding gauge coupling. As only the SM leptons and right-handed neutrinos of muon and tauon flavors are charged fermions under such a gauge symmetry, the neutral-current interactions of these fermions with $Z^\prime_\mu$ read
\begin{eqnarray}
    {\cal L}^\prime_{\rm NC} = g^{}_{Z^\prime} \left[\left(\overline{\ell^{}_{\mu {\rm L}}} \gamma^\mu \ell^{}_{\mu {\rm L}} - \overline{\ell^{}_{\tau {\rm L}}} \gamma^\mu \ell^{}_{\tau {\rm L}}\right) + \left( \overline{\mu^{}_{\rm R}} \gamma^\mu \mu^{}_{\rm R} - \overline{\tau^{}_{\rm R}} \gamma^\mu \tau^{}_{\rm R} \right) + \left( \overline{\nu^{}_{\mu{\rm R}}} \gamma^\mu \nu^{}_{{\mu}\rm R} - \overline{\nu^{}_{{\tau}\rm R}} \gamma^\mu \nu^{}_{{\tau}\rm R} \right)\right]Z^\prime_\mu \; , \quad
    \label{eq:newNC}
\end{eqnarray}
where the interaction terms in the first two round brackets on the right-hand side will contribute to the muon $(g-2)$, as we shall see later.

\begin{table}[t!]
\centering
\caption{The charge assignments of the relevant fields under the ${\rm SU}(2)^{}_{\rm L}\otimes {\rm U}(1)^{}_{\rm Y}\otimes {\rm U}(1)^{}_{L^{}_\mu - L^{}_\tau}$ gauge symmetry in the extended type-(I+II) seesaw model.}
\vspace{0.5cm}
\begin{tabular}{ccccccccc}
\toprule
		& $\ell^{}_{e{\rm L}}, \ell^{}_{\mu{\rm L}}, \ell^{}_{\tau{\rm L}}$ & $e^{}_{\rm R}, \mu^{}_{\rm R}, \tau^{}_{\rm R}$ & $\nu^{}_{e{\rm R}}, \nu^{}_{\mu{\rm R}}, \nu^{}_{\tau{\rm R}}$ & $S$ & $H$ & $\Delta$ \\
\midrule
		${\rm SU}(2)^{}_{\rm L}\otimes U(1)^{}_{\rm Y}$ & ({\bf 2}, $-1$) & ({\bf 1}, $-2$) & ({\bf 1}, 0) & ({\bf 1}, 0) & ({\bf 2}, +1) & ({\bf 3}, $-2$)  \\
		$U(1)^{}_{L^{}_\mu - L^{}_\tau}$ & $0$, $+1$, $-1$ & $0$, $+1$, $-1$ & $0$, $+1$, $-1$ & $+1$ & $0$ & $-1$  \\
\bottomrule
\end{tabular}
\label{Table: model}
\end{table}

Second, the scalar potential of the model resembles much that of the type-II seesaw model~\cite{Arhrib:2011uy}, where only one Higgs triplet is introduced into the SM. With an extra singlet scalar $S$, one can immediately write down the gauge-invariant scalar potential
\begin{eqnarray}
    V(S, H, \Delta) &=& -\mu^2_H H^\dagger H + \lambda^{}_H (H^\dagger H)^2 - \mu^2_S S^\dagger S + \lambda^{}_S (S^\dagger S)^2 + \frac{1}{2} M^2_\Delta {\rm Tr}(\Delta^\dagger \Delta) \nonumber \\
     &~& + \left(\lambda^{}_{1} S H^{\rm T} {\rm i}\sigma^{}_2 \Delta H + {\rm h.c.} \right) + \frac{\lambda^{}_2}{2} (H^\dagger H) {\rm Tr}(\Delta^\dagger \Delta) + \frac{\lambda^{}_3}{4} (H^\dagger \sigma^{}_i H) {\rm Tr}(\Delta^\dagger \sigma^{}_i \Delta) \nonumber \\
     &~& + \frac{\lambda^{}_4}{2} (S^\dagger S) {\rm Tr}(\Delta^\dagger \Delta) + \frac{\lambda^{}_5}{2} (S^\dagger S) (H^\dagger H) + \frac{\lambda^{}_6}{4} {\rm Tr}\left[ (\Delta^\dagger \Delta)^2\right] + \frac{\lambda^{}_7}{4} \left[{\rm Tr}(\Delta^\dagger \Delta)\right]^2 \; ,
     \label{eq:Vsdt}
    \end{eqnarray}
where the coupling constant $\lambda^{}_1$ is in general complex. It should be noticed that the coefficient of the trilinear coupling term of the SM Higgs doublet $H$ and the Higgs triplet $\Delta$ is of mass dimension and small in the type-II seesaw model, whereas it is now given by $\lambda^{}_1 \langle S \rangle = \lambda^{}_1 v^{}_S$ after the singlet scalar acquires its vacuum expectation value (vev). Instead of a complete analysis of the vacuum structure based on the scalar potential $V(S, H, \Delta)$, we make some reasonable approximations and recapitulate the main features of the spontaneous symmetry breaking in our model.
\begin{itemize}
\item The general scalar potential in Eq.~(\ref{eq:Vsdt}) can be greatly simplified if some quartic couplings (e.g., $\lambda^{}_3$, $\lambda^{}_4$, $\lambda^{}_5$ and $\lambda^{}_7$) are set to zero. In this case, the physical triplet scalars $H^{\pm \pm}$, $H^\pm$, $H^0$ and $A^0$ are degenerate in mass, namely, $M^{}_{H^{\pm\pm}} = M^{}_{H^\pm} = M^{}_{H^0} = M^{}_{A^0} = M^{}_\Delta$. Meanwhile, the vev's of the neutral scalar bosons are approximately given by $v^{}_S \approx \sqrt{\mu^2_S/\lambda^{}_S}$, $v^{}_H \approx \sqrt{\mu^2_H/\lambda^{}_H}$, and $v^{}_\Delta \approx \lambda^{}_1 v^{}_S v^2_H/M^2_\Delta$. Note that the coupling constant $\lambda^{}_1$ can be made real by redefining the phases of scalar fields $S$, $H$ and $\Delta$.

\item The singlet scalar field $S$ is only charged under the ${\rm U}(1)^{}_{L^{}_\mu - L^{}_\tau}$ gauge symmetry, so the mass of the gauge boson $Z^\prime$ is given by $M^{}_{Z^\prime} = g^{}_{Z^\prime} v^{}_S/2$. As the Higgs triplet carries the charges of all three gauge symmetries, it will induce the mass mixing among three neutral gauge bosons $Z$, $Z^\prime$ and $\gamma$ after the spontaneous symmetry breaking. However, since the vev of the triplet Higgs is constrained severely by the precision measurement of the $\rho$ parameter~\cite{PDG2020}, namely, $\rho \equiv M^2_W/(M^2_Z \cos^2 \theta^{}_{\rm W}) \approx (v^2_H + 2v^2_\Delta)/(v^2_H + 4 v^2_\Delta) = 1.00038 \pm 0.00020$, from which one can set an upper bound $v^{}_\Delta \lesssim 2.58~{\rm GeV}$ at the $3\sigma$ level for $v^{}_H \approx 246~{\rm GeV}$, one expects that $M^{}_{Z^\prime}$ is mainly determined by the gauge coupling $g^{}_{Z^\prime}$ and the vev $v^{}_S$ of the singlet scalar. In our case, $v^{}_\Delta$ will also contribute to light neutrino masses, and thus its magnitude will be further required to be below $1~{\rm eV}$. Consequently, the dominant decay channel of the doubly-charged and singly-charged Higgs bosons is purely leptonic.
\end{itemize}

Third, the gauge-invariant Yukawa interactions and mass terms for leptons turn out to be
    \begin{eqnarray}
    {\cal L}^{}_{\rm lepton} &=& - y^\alpha_l \overline{\ell^{}_{\alpha {\rm L}}} \alpha^{}_{\rm R} H - \frac{1}{2} y^{}_\Delta \left( \overline{\ell^{}_{e{\rm L}}} \Delta {\rm i}\sigma^{}_2 \ell^{\rm C}_{\tau {\rm L}} + \overline{\ell^{}_{\tau{\rm L}}} \Delta {\rm i}\sigma^{}_2 \ell^{\rm C}_{e {\rm L}}\right) - y^\alpha_\nu \overline{\ell^{}_{\alpha {\rm L}}} \tilde{H} \nu^{}_{\alpha {\rm R}} \nonumber \\
    &~& - \frac{1}{2} y^{e\tau}_{S} \left( \overline{\nu^{\rm C}_{e{\rm R}}} \nu^{}_{\tau {\rm R}} + \overline{\nu^{\rm C}_{\tau{\rm R}}} \nu^{}_{e {\rm R}} \right) S - \frac{1}{2} y^{e\mu}_{S} \left( \overline{\nu^{\rm C}_{e{\rm R}}} \nu^{}_{\mu {\rm R}} + \overline{\nu^{\rm C}_{\mu{\rm R}}} \nu^{}_{e {\rm R}} \right) S^\dagger \nonumber \\
    &~& - \frac{1}{2} \left[ m^{ee}_{\rm R} \overline{\nu^{\rm C}_{e {\rm R}}} \nu^{}_{e {\rm R}} + m^{\mu\tau}_{\rm R} \left(\overline{\nu^{\rm C}_{\mu {\rm R}}} \nu^{}_{\tau {\rm R}}  + \overline{\nu^{\rm C}_{\tau {\rm R}}} \nu^{}_{\mu {\rm R}} \right)\right]  + {\rm h.c.} \; ,
    \label{eq:Yukawalep}
    \end{eqnarray}
    which after the spontaneous symmetry breaking with $\langle H \rangle = v^{}_H/\sqrt{2}$, $\langle \Delta \rangle = v^{}_\Delta$ and $\langle S \rangle = v^{}_S/\sqrt{2}$ leads to the lepton mass matrices
    \begin{eqnarray}
    M^{}_l = \frac{v^{}_H}{\sqrt{2}} \left( \begin{matrix} y^e_l & 0 & 0 \cr 0 & y^\mu_l & 0 \cr 0 & 0 & y^\tau_l \end{matrix} \right) \; , \quad M^{}_{\rm D} =  \frac{v^{}_H}{\sqrt{2}} \left( \begin{matrix} y^e_\nu & 0 & 0 \cr 0 & y^\mu_\nu & 0 \cr 0 & 0 & y^\tau_\nu \end{matrix} \right) \; ,
    \label{eq:MlMD}
    \end{eqnarray}
    and
    \begin{eqnarray}
     M^{}_{\rm L} =  \left( \begin{matrix} 0 & 0 & y^{}_\Delta v^{}_\Delta \cr 0 & 0 & 0 \cr y^{}_\Delta v^{}_\Delta & 0 & 0 \end{matrix} \right) \; , \quad M^{}_{\rm R} = \left( \begin{matrix} m^{ee}_{\rm R} & y^{e\mu}_S v^{}_S/\sqrt{2} & y^{e\tau}_S v^{}_S/\sqrt{2} \cr y^{e\mu}_S v^{}_S/\sqrt{2} & 0 & m^{\mu\tau}_{\rm R} \cr y^{e\tau}_S v^{}_S/\sqrt{2} & m^{\mu\tau}_{\rm R} & 0 \end{matrix}\right) \; .
     \label{eq:MLMR}
    \end{eqnarray}

Before going into the details of neutrino masses and leptonic flavor mixing, we examine the relevant model parameters in the leptonic sector. First, as indicated in Eq.~(\ref{eq:Yukawalep}), the charged-lepton Yukawa coupling matrix $Y^{}_l \equiv {\rm Diag}\{y^e_l, y^\mu_l, y^\tau_l \}$ is diagonal, and one can always choose the coupling constants $y^\alpha_l$ (for $\alpha = e, \mu, \tau$) to be real by redefining the phases of right-handed charged-lepton fields. In this case, the charged-lepton masses are simply given by $m^{}_\alpha = y^\alpha_l v^{}_H/\sqrt{2}$ (for $\alpha = e, \mu, \tau$). Second, without loss of generality, we can also make the Dirac neutrino Yukawa coupling constants $y^\alpha_\nu$ (for $\alpha = e, \mu, \tau$) all real by absorbing their phases into the left-handed lepton doublets. In addition, the overall phase of $y^{}_\Delta$ can be removed by rephasing the Higgs triplet $\Delta$, so the type-II Majorana neutrino mass matrix $M^{}_{\rm L}$ is symmetric and real. Finally, one can eliminate the phases of the mass parameters $m^{ee}_{\rm R}$ and $m^{\mu\tau}_{\rm R}$ by utilizing the freedom of redefining the phases of right-handed neutrinos. However, it is impossible to choose both $y^{e\mu}_S$ and $y^{e\tau}_S$ to be real, even with the phase redefinition of the singlet scalar field $S$. For illustration, we take $y^{e\mu}_S$ to be complex and all the other model parameters to be real and positive in the following discussions. The other case with a complex $y^{e\tau}_S$ but a real $y^{e\mu}_S$ can be similarly studied.

\section{Neutrino Masses and Flavor Mixing}
\label{sec:mass}

As the Dirac neutrino mass matrix $M^{}_{\rm D} = {\rm Diag}\{d^{}_e, d^{}_\mu, d^{}_\tau\}$ with $d^{}_\alpha \equiv y^\alpha_\nu v^{}_H/\sqrt{2}$ is simply diagonal, the leptonic flavor mixing arises solely from the flavor structure of $M^{}_{\rm R}$ and its interplay with the contribution from the Higgs triplet. Without the Higgs triplet, the phenomenological implications for neutrino masses and leptonic flavor mixing have been thoroughly studied in a numerical way in the literature~\cite{Biswas:2016yan}. In the assumption of $y^{e\tau}_S \ll |y^{e\mu}_S|$, we have found that it is possible to perform an analytical calculation, which will manifest the direct correlation between model parameters and physical observables.

Ignoring the element $y^{e\tau}_S v^{}_S/\sqrt{2}$ in the Majorana mass matrix of right-handed neutrinos $M^{}_{\rm R}$, we observe that it can be exactly diagonalized via the orthogonal transformation $U^\dagger M^{}_{\rm R} U^* = \widehat{M}^{}_{\rm R} = {\rm Diag}\{M^{}_1, M^{}_2, M^{}_3\}$, where $M^{}_i$ (for $i = 1, 2, 3$) are heavy Majorana neutrino masses and related to the real and positive parameters by~\cite{Xing:2004xu, Xing:2002sb}
\begin{eqnarray}
m^e_{\rm R} &=& M^{}_3 (1 - y^{}_{\rm R} + x^{}_{\rm R} y^{}_{\rm R}) \; ,
\label{eq:A}
\\
\frac{|y^{e\mu}_S| v^{}_S}{\sqrt{2}} &=& M^{}_3 \left[ \frac{y^{}_{\rm R} (1 - x^{}_{\rm R}) (1 - y^{}_{\rm R}) (1 + x^{}_{\rm R} y^{}_{\rm R})}{1 - y^{}_{\rm R} + x^{}_{\rm R} y^{}_{\rm R}} \right]^{1/2} \; ,
\label{eq:B}
\\
m^{\mu\tau}_{\rm R} &=& M^{}_3 \left( \frac{x^{}_{\rm R} y^2_{\rm R}}{1 - y^{}_{\rm R} + x^{}_{\rm R} y^{}_{\rm R}} \right)^{1/2} \; ,
\label{eq:C}
\end{eqnarray}
where the mass ratios $x^{}_{\rm R} \equiv M^{}_1/M^{}_2$ and $y^{}_{\rm R} \equiv M^{}_2/M^{}_3$ have been defined with both $x^{}_{\rm R}$ and $y^{}_{\rm R}$ being in the range $(0, 1)$. The nine elements $U^{}_{\alpha i}$ (for $\alpha = e, \mu, \tau$ and $i = 1, 2, 3$) of the unitary matrix $U$ are explicitly given by~\cite{Xing:2004xu}
\begin{eqnarray}
U^{}_{e1} &=& - \left[ \frac{x^{}_{\rm R}y^{}_{\rm R}(1-x^{}_{\rm R})(1+x^{}_{\rm R}y^{}_{\rm R})}{(1+x^{}_{\rm R})(1-x^{}_{\rm R}y^{}_{\rm R})(1-y^{}_{\rm R}+x^{}_{\rm R}y^{}_{\rm R})} \right]^{1/2} \; , \nonumber \\
U^{}_{e2} &=& -{\rm i} \left[ \frac{y^{}_{\rm R} (1 - x^{}_{\rm R}) (1 - y^{}_{\rm R})}{(1 + x^{}_{\rm R}) (1 + y^{}_{\rm R}) (1 - y^{}_{\rm R} + x^{}_{\rm R} y^{}_{\rm R})} \right]^{1/2} \; , \nonumber \\
U^{}_{e3} &=& + \left[ \frac{(1 - y^{}_{\rm R}) (1 + x^{}_{\rm R} y^{}_{\rm R})}{(1 - x^{}_{\rm R} y^{}_{\rm R})(1 + y^{}_{\rm R})(1 - y^{}_{\rm R} + x^{}_{\rm R} y^{}_{\rm R})} \right]^{1/2} \; , \nonumber \\
U^{}_{\mu 1} &=& + e^{{\rm i}\phi} \left[ \frac{x^{}_{\rm R}(1 - y^{}_{\rm R})}{(1 + x^{}_{\rm R}) (1 - x^{}_{\rm R} y^{}_{\rm R})} \right]^{1/2} \; , \nonumber \\
U^{}_{\mu 2} &=& + {\rm i} e^{{\rm i}\phi} \left[ \frac{1 + x^{}_{\rm R} y^{}_{\rm R}}{(1 + x^{}_{\rm R}) (1 + y^{}_{\rm R})} \right]^{1/2} \; , \nonumber \\
U^{}_{\mu 3} &=& + e^{{\rm i}\phi} \left[ \frac{y^{}_{\rm R} (1 - x^{}_{\rm R})}{(1 - x^{}_{\rm R} y^{}_{\rm R})(1 + y^{}_{\rm R})} \right]^{1/2} \; , \nonumber \\
U^{}_{\tau 1} &=& + e^{-{\rm i}\phi} \left[ \frac{1 - y^{}_{\rm R}}{(1 + x^{}_{\rm R})(1 - x^{}_{\rm R} y^{}_{\rm R})(1 - y^{}_{\rm R} + x^{}_{\rm R} y^{}_{\rm R})} \right]^{1/2} \; , \nonumber \\
U^{}_{\tau 2} &=& -{\rm i} e^{-{\rm i}\phi}  \left[ \frac{x^{}_{\rm R}(1 + x^{}_{\rm R} y^{}_{\rm R})}{(1 + x^{}_{\rm R})(1 + y^{}_{\rm R})(1 - y^{}_{\rm R} + x^{}_{\rm R} y^{}_{\rm R})} \right]^{1/2} \; , \nonumber \\
U^{}_{\tau 3} &=& + e^{-{\rm i}\phi} \left[ \frac{x^{}_{\rm R}y^3_{\rm R}(1 - x^{}_{\rm R})}{(1 - x^{}_{\rm R} y^{}_{\rm R}) (1 + y^{}_{\rm R}) (1 - y^{}_{\rm R} + x^{}_{\rm R} y^{}_{\rm R})} \right]^{1/2} \; ,
\label{eq:U}
\end{eqnarray}
where the phase $\phi$ has been defined as $y^{e\mu}_S \equiv |y^{e\mu}_S| e^{{\rm i}\phi}$. As one can see from the above discussions, three heavy Majorana neutrino masses $M^{}_i$ (for $i = 1, 2, 3$) or equivalently $\{x^{}_{\rm R}, y^{}_{\rm R}, M^{}_3\}$ can be implemented to replace the model parameters $m^e_{\rm R}$, $|y^{e\mu}_S| v^{}_S$ and $m^{\mu\tau}_{\rm R}$. The effective Majorana mass matrix of three light neutrinos can be obtained from the type-(I+II) seesaw formula, i.e.,
\begin{eqnarray}
M^{}_\nu \approx M^{}_{\rm L} - M^{}_{\rm D} M^{-1}_{\rm R} M^{\rm T}_{\rm D} \; ,
\label{eq:Mnutype2}
\end{eqnarray}
from which the explicit expressions of six independent matrix elements are found to be
\begin{eqnarray}
\left(M^{}_\nu\right)^{}_{ee} &=& - \frac{d^2_e}{M^{}_3} \cdot \frac{1}{1 - y^{}_{\rm R} + x^{}_{\rm R} y^{}_{\rm R}} \; , \nonumber \\ \left(M^{}_\nu\right)^{}_{\mu\mu} &=& 0 \; , \nonumber \\ \left(M^{}_\nu\right)^{}_{\tau\tau} &=& -\frac{d^2_\tau}{M^{}_3} \cdot \frac{(1 - x^{}_{\rm R}) (1 - y^{}_{\rm R}) (1 + x^{}_{\rm R} y^{}_{\rm R})}{x^{}_{\rm R} y^{}_{\rm R} (1 - y^{}_{\rm R} + x^{}_{\rm R} y^{}_{\rm R} )} \cdot e^{2{\rm i} \phi} \; ,
\label{eq:Mnudiag}
\end{eqnarray}
and
\begin{eqnarray}
\left(M^{}_\nu\right)^{}_{e\mu} &=& 0 \; , \nonumber \\ \left(M^{}_\nu\right)^{}_{e\tau} &=& y^{}_\Delta v^{}_\Delta + \frac{d^{}_e d^{}_\tau}{M^{}_3} \cdot \frac{1}{1 - y^{}_{\rm R} + x^{}_{\rm R} y^{}_{\rm R}} \cdot \left[ \frac{(1 - x^{}_{\rm R}) (1 - y^{}_{\rm R}) (1 + x^{}_{\rm R} y^{}_{\rm R})}{x^{}_{\rm R} y^{}_{\rm R}} \right]^{1/2} \cdot e^{{\rm i}\phi}\; , \nonumber \\
\left(M^{}_\nu\right)^{}_{\mu\tau} &=& -\frac{d^{}_\mu d^{}_\tau}{M^{}_3} \cdot \left( \frac{x^{}_{\rm R} y^2_{\rm R}}{1 - y^{}_{\rm R} + x^{}_{\rm R} y^{}_{\rm R}} \right)^{1/2} \; .
\label{eq:Mnuoff}
\end{eqnarray}
One can immediately realize that the effective Majorana neutrino mass matrix $M^{}_\nu$ with $(M^{}_\nu)^{}_{e\mu} = (M^{}_\nu)^{}_{\mu\mu} = 0$ takes the form of the two-zero texture ${\bf B}^{}_3$, for which the implications for neutrino masses and leptonic flavor mixing have carefully been examined in the previous literature~\cite{Fritzsch:2011qv, Zhou:2015qua}. Some helpful comments are in order.
\begin{itemize}
\item In the hierarchical limit of heavy Majorana neutrino masses, i.e., $M^{}_1 \ll M^{}_2 \ll M^{}_3$ or equivalently $x^{}_{\rm R}, y^{}_{\rm R} \ll 1$, the expressions of nonzero neutrino mass matrix elements can be greatly simplified. More explicitly, we have $\left(M^{}_\nu\right)^{}_{ee} = - d^2_e/M^{}_3$, $\left(M^{}_\nu\right)^{}_{\tau\tau} = - d^2_\tau e^{2{\rm i}\phi}/(M^{}_3 x^{}_{\rm R} y^{}_{\rm R})$, $\left(M^{}_\nu\right)^{}_{e\tau} = y^{}_\Delta v^{}_\Delta + d^{}_e d^{}_\tau e^{{\rm i}\phi}/(M^{}_3 \sqrt{x^{}_{\rm R} y^{}_{\rm R}})$, and $\left(M^{}_\nu\right)^{}_{\mu\tau} = - d^{}_\mu d^{}_\tau \sqrt{x^{}_{\rm R}} y^{}_{\rm R}/M^{}_3$. Since the charged-lepton mass matrix $M^{}_l = {\rm Diag}\{m^{}_e, m^{}_\mu, m^{}_\tau\}$ is diagonal, the effective neutrino mass matrix $M^{}_\nu$ can directly be reconstructed from the leptonic mixing matrix $V$ and three neutrino masses $\widehat{M}^{}_\nu \equiv {\rm Diag}\{m^{}_1, m^{}_2, m^{}_3\}$ via $M^{}_\nu = V \widehat{M}^{}_\nu V^{\rm T}$. With the precision measurements of neutrino masses and flavor mixing parameters, one can extract very useful information about the model parameters from Eqs.~(\ref{eq:Mnudiag}) and (\ref{eq:Mnuoff}).

\item For the two-zero texture ${\bf B}^{}_3$ of the neutrino mass matrix $M^{}_\nu$, the zero elements results in serious constraints on neutrino masses and flavor mixing parameters. Given $(M^{}_\nu)^{}_{e\mu} = (M^{}_\nu)^{}_{\mu\mu} = 0$, we find~\cite{Xing:2002ta, Xing:2002ap, Fritzsch:2011qv, Zhou:2015qua}
    \begin{eqnarray}
    \frac{\lambda^{}_1}{\lambda^{}_2}  &=& - \tan \theta^{}_{23} \cdot \frac{\sin \theta^{}_{12} \sin \theta^{}_{23} + \cos \theta^{}_{12} \cos \theta^{}_{23} \sin \theta^{}_{13} e^{-{\rm i}\delta}}{\sin \theta^{}_{12} \cos \theta^{}_{23} - \cos \theta^{}_{12} \sin \theta^{}_{23} \sin \theta^{}_{13} e^{+{\rm i}\delta}} \; , \nonumber \\
    \frac{\lambda^{}_2}{\lambda^{}_3} &=& - \tan \theta^{}_{23} \cdot \frac{\cos \theta^{}_{12} \sin \theta^{}_{23} - \sin \theta^{}_{12} \cos \theta^{}_{23} \sin \theta^{}_{13} e^{-{\rm i}\delta}}{\cos \theta^{}_{12} \cos \theta^{}_{23} + \sin \theta^{}_{12} \sin \theta^{}_{23} \sin \theta^{}_{13} e^{+{\rm i}\delta}} \; ,
    \label{eq:lambda}
    \end{eqnarray}
    where $\lambda^{}_1 \equiv m^{}_1 e^{2{\rm i}\rho}$, $\lambda^{}_2 \equiv m^{}_2 e^{2{\rm i}\sigma}$ and $\lambda^{}_3 \equiv m^{}_3$ with $\rho$ and $\sigma$ being two Majorana-type CP-violating phases. Notice that the standard parametrization of the leptonic flavor mixing matrix $V$ has been adopted~\cite{PDG2020}, where $\{\theta^{}_{12}, \theta^{}_{13}, \theta^{}_{23}\}$ are three flavor mixing angles and $\delta$ is the Dirac-type CP-violating phase. In the leading-order approximation, we ignore the terms proportional to $\sin\theta^{}_{13} \approx 0.15$ in Eq.~(\ref{eq:lambda}) and thus observe that $m^{}_1/m^{}_2 \approx m^{}_2/m^{}_3 \approx \tan^2 \theta^{}_{23}$. Therefore, if neutrino mass ordering is normal (i.e., $m^{}_1 < m^{}_2 < m^{}_3$), then $\theta^{}_{23} < 45^\circ$ should hold. Although the best-fit value $\theta^{}_{23} \approx 49^\circ$ is found by the global-fit analysis of current neutrino oscillation data~\cite{Esteban:2020cvm}, the values $39.6^\circ \lesssim \theta^{}_{23} \lesssim 51.8^\circ$ are still allowed at the $3\sigma$ level. If the neutrino mass ordering is inverted (i.e., $m^{}_3 < m^{}_1 < m^{}_2$), then $\theta^{}_{23} > 45^\circ$ is required. Such a correlation between neutrino mass ordering and the octant of $\theta^{}_{23}$ will definitely be tested in future neutrino oscillation experiments. At the next-to-leading order, one can determine the Dirac CP-violating phase via
    \begin{eqnarray}
     \cos \delta \approx - \frac{\Delta m^2_{21}}{\Delta m^2_{31}} \cdot \frac{\sin2\theta^{}_{12} \cot^2 \theta^{}_{23}}{2\sin \theta^{}_{13} \tan 2\theta^{}_{23}} \; ,
     \label{eq:cosd}
     \end{eqnarray}
     where $\Delta m^2_{ij} \equiv m^2_i - m^2_j$ (for $ij = 21, 31$) are two neutrino mass-squared differences. In the case of normal neutrino mass ordering (i.e., $\Delta m^2_{21} > 0$ and $\Delta m^2_{31} > 0$), we have $\cos\delta < 0$ because of $\tan 2\theta^{}_{23} > 0$ for $\theta^{}_{23} < 45^\circ$. It is straightforward to figure out that $\cos\delta < 0$ is also valid in the case of inverted neutrino mass ordering.
\end{itemize}

In summary, in the limit of $y^{e\tau}_S \ll |y^{e\mu}_S|$, one can exactly diagonalize the right-handed neutrino mass matrix $M^{}_{\rm R}$, and the effective Majorana neutrino mass matrix $M^{}_\nu \approx M^{}_{\rm L} - M^{}_{\rm D} M^{-1}_{\rm R} M^{\rm T}_{\rm D}$ turns out to be of the form of two-zero texture ${\bf B}^{}_3$. At the $3\sigma$ level, the predictions for neutrino masses and flavor mixing parameters are compatible with current neutrino oscillation data. As there exists a direct connection between neutrino mass ordering and the octant of $\theta^{}_{23}$, the predictions will be well testable in the next-generation neutrino oscillation experiments.

\section{Anomalous Magnetic Moments}
\label{sec:mag}
\begin{figure}[t!]
\begin{center}	
\hspace{0.5cm}
\includegraphics[width=0.95\textwidth]{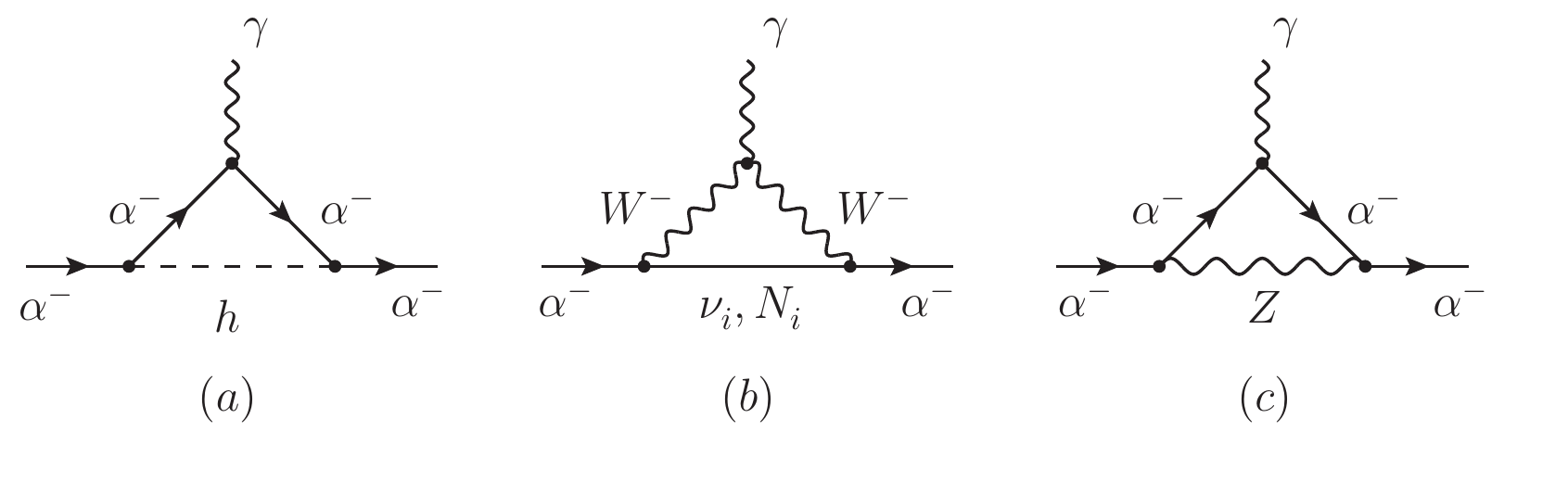}
\end{center}
\vspace{-1.2cm}
\caption{The Feynman diagrams for the contributions to the anomalous magnetic moments of the charged lepton $\alpha^-$ (for $\alpha = e, \mu, \tau$) in the type-(I+II) seesaw model: (a) the SM Higgs boson $h$; (b) the SM charged gauge boson $W^-$ and three light ($\nu^{}_i$) or heavy ($N^{}_i$) Majorana neutrinos; (c) the SM neutral gauge boson $Z$.}
\label{fig:alphag} 
\end{figure}

Apart from neutrino oscillation parameters, the contributions from the new particles beyond the SM to the anomalous electron and muon magnetic moments can be calculated. To this end, it is helpful to specify the new particles appearing in the type-(I+II) seesaw model under consideration and their interactions with charged leptons.

In addition to the electromagnetic interactions, there are three types of contributions to the charged-lepton magnetic moment in the SM~\cite{Jackiw:1972jz, Bars:1972pe, Fujikawa:1972fe, Bardeen:1972vi}, for which the Feynman diagrams are listed in Fig.~\ref{fig:alphag}. Since the electroweak contributions are well known, we focus only on the new-physics ones. Due to the ${\rm U}(1)^{}_{L^{}_\mu - L^{}_\tau}$ symmetry, the Higgs triplet $\Delta$ is coupled only to the electron and tauon flavors in a flavor-changing way. As a consequence, it contributes to the anomalous magnetic moments of electron and tauon, which will be considered later on. For the flavor-universal contributions, only the Feynman diagram in Fig.~\ref{fig:alphag}(b) involving massive Majorana neutrinos matters. This is actually the case in the type-I seesaw model. In the basis of left-handed neutrino states $\nu^{}_{\alpha {\rm L}}$ and $\nu^{\rm C}_{\alpha {\rm R}}$, the overall $6\times 6$ mass matrix of neutrinos can be diagonalized via~\cite{Xing:2019vks}
\begin{eqnarray}
\left( \begin{matrix} M^{}_{\rm L} & M^{}_{\rm D} \cr M^{\rm T}_{\rm D} & M^{}_{\rm R}\end{matrix} \right) = \left( \begin{matrix} {\cal V} & {\cal R} \cr {\cal S} & {\cal U}\end{matrix} \right) \left( \begin{matrix} \widehat{M}^{}_\nu & 0 \cr 0 & \widehat{M}^{}_{\rm R} \end{matrix} \right) \left( \begin{matrix} {\cal V} & {\cal R} \cr {\cal S} & {\cal U}\end{matrix} \right)^{\rm T} \; ,
\label{eq:six}
\end{eqnarray}
where ${\cal V}$, ${\cal R}$, ${\cal S}$ and ${\cal U}$ are $3\times 3$ non-unitary matrices satisfying the unitarity conditions ${\cal V}{\cal V}^\dagger + {\cal R}{\cal R}^\dagger = {\cal S} {\cal S}^\dagger + {\cal U} {\cal U}^\dagger = {\bf 1}$ and ${\cal V}{\cal S}^\dagger + {\cal R}{\cal U}^\dagger = {\cal S} {\cal V}^\dagger + {\cal U} {\cal R}^\dagger = {\bf 0}$. For ${\cal O}(M^{}_{\rm D}) \ll {\cal O}(M^{}_{\rm R})$ as in the canonical type-I seesaw model, one can relate the non-unitary matrices ${\cal V}$ and ${\cal R}$ to the unitary matrices $V$ and $U$ introduced in the previous section by ${\cal V} \approx ({\bf 1} - {\cal R} {\cal R}^\dagger/2)V$ and ${\cal R} \approx M^{}_{\rm D} U^* \widehat{M}^{-1}$. In the mass basis, the leptonic charged-current interactions read
\begin{eqnarray}
{\cal L}^{}_{\rm CC} = \frac{g}{\sqrt{2}} \overline{\alpha^{}_{\rm L}} \gamma^\mu ({\cal V}^{}_{\alpha i} \nu^{}_i + {\cal R}^{}_{\alpha i} N^{}_i) W^-_\mu + {\rm h.c.} \; ,
\label{eq:CC}
\end{eqnarray}
where $\nu^{}_i$ and $N^{}_i$ (for $i = 1, 2, 3$) stand for the mass eigenfields of three light and heavy Majorana neutrinos, respectively. With the charged-current interactions in Eq.~(\ref{eq:CC}) and the general formulas derived in Refs.~\cite{Leveille:1977rc, Moore:1984eg}, we can obtain the contributions from light and heavy Majorana neutrinos to the anomalous magnetic moments of three charged leptons
\begin{eqnarray}
a^{W}_\alpha &=& \frac{G^{}_{\rm F} m^2_\alpha}{4 \sqrt{2} \pi^2} \left\{ \sum^3_{i = 1} {\cal V}^{}_{\alpha i} {\cal V}^*_{\alpha i} G\left(\frac{m^2_i}{M^2_W}\right) + \sum^3_{i = 1} {\cal R}^{}_{\alpha i} {\cal R}^*_{\alpha i} G\left(\frac{M^2_i}{M^2_W}\right) \right\} \; ,
\label{eq:aW}
\end{eqnarray}
where the loop function is defined as
\begin{eqnarray}
G(z) \equiv \int^1_0 {\rm d}x \frac{2x^2 (1+x) + z(2x - 3x^2 + x^3) }{ x + z (1 - x)} = \frac{10 - 43 z + 78 z^2 - 49 z^3 + 4 z^4 + 18 z^3 \ln z}{6(1-z)^4} \; . \quad
\label{eq:loopf}
\end{eqnarray}
It is straightforward to verify $G(z) = 5/3 - z/2 + {\cal O}(z^2)$ in the limit of $z \ll 1$ and $G(z) = 2/3 - (11 - 6\ln z)/(2z) + {\cal O}(z^{-2})$ in the limit of $z \gg 1$. The SM value can easily be obtained from Eq.~(\ref{eq:aW}) by setting ${\cal R} \to {\bf 0}$, ${\cal V}{\cal V}^\dagger \to {\bf 1}$ and vanishing neutrino masses. After subtracting the SM contribution, we get the contributions from massive Majorana neutrinos, i.e.,
\begin{eqnarray}
\Delta a^{W}_\alpha = \frac{G^{}_{\rm F} m^2_\alpha}{4 \sqrt{2} \pi^2} \left\{ \sum^3_{i = 1} {\cal R}^{}_{\alpha i} {\cal R}^*_{\alpha i} \left[ G\left(\frac{M^2_i}{M^2_W}\right) - \frac{5}{3}\right] - \frac{1}{2} \sum^3_{i = 1} {\cal V}^{}_{\alpha i} {\cal V}^*_{\alpha i} \frac{m^2_i}{M^2_W}\right\} \; .
\label{eq:DeltaaW}
\end{eqnarray}
It is worthwhile to point out that the loop function in Eq.~(\ref{eq:loopf}) and the final result in Eq.~(\ref{eq:DeltaaW}) are different from those in Ref.~\cite{Biggio:2008in} for the type-I seesaw model. However, the conclusions from Ref.~\cite{Biggio:2008in} remain to be valid, namely, the value of $[G(M^2_i/M^2_W) - 5/3]$ is always negative and thus $\Delta a^W_\mu$ has the wrong sign. For illustration, if we take $M^{}_i = 1~{\rm TeV}$ (i.e., the limit of $M^2_i/M^2_W \gg 1$ holds) and ignore the tiny neutrino masses $m^{}_i$, then
\begin{eqnarray}
\Delta a^W_\mu \approx -\frac{G^{}_{\rm F} m^2_\mu}{4 \sqrt{2} \pi^2} \left({\cal R} {\cal R}^\dagger\right)^{}_{\mu\mu} \approx - 0.1\times 10^{-11} \; ,
\label{eq:DeltaaWn}
\end{eqnarray}
for $({\cal R}{\cal R}^\dagger)^{}_{\mu\mu} \approx 4.4\times 10^{-4}$ saturating the upper bounds from electroweak precision data~\cite{Fernandez-Martinez:2016lgt}. Compared to the experimental observation in Eq.~(\ref{eq:Deltaa}), the correction from $\Delta a^W_\mu$ is negligible.
\begin{figure}[t!]
\begin{center}		
\hspace{0.5cm}
\includegraphics[width=0.95\textwidth]{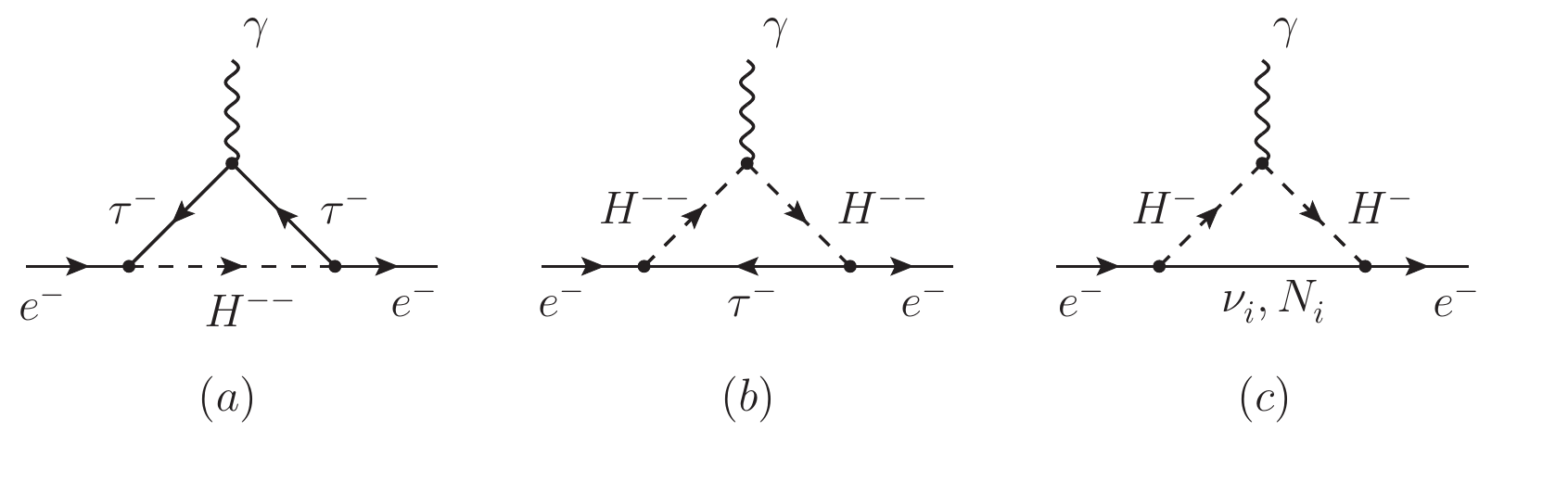}
\end{center}
\vspace{-1.2cm}
\caption{The Feynman diagrams for the contributions to the anomalous electron magnetic moment in the type-(I+II) seesaw model: (a) and (b) from the doubly-charged Higgs boson $H^{--}$; and (c) from the singly-charged Higgs boson $H^-$.}
\label{fig:electrong} 
\end{figure}

Next, we turn to the impact of the triplet Higgs on the anomalous magnetic moment of electron $a^{}_e \equiv (g^{}_e - 2)/2$. The latest measurement of the fine structure constant leads to a reevaluation of $a^{}_e$ and a $1.6\sigma$ discrepancy with the theoretical prediction has been found~\cite{Morel:2020dww}
\begin{eqnarray}
\Delta a^{}_e \equiv a^{\rm exp}_e - a^{\rm SM}_e = 0.48(30)\times 10^{-12} \; ,
\label{eq:Deltaae}
\end{eqnarray}
whereas the discrepancy from the caesium recoil measurements is even larger $\Delta a^{}_e \equiv a^{\rm exp}_e - a^{\rm SM}_e = -0.88(36)\times 10^{-12}$ with an opposite sign~\cite{Parker:2018vye}. At the $2\sigma$ level, either positive or negative sign of $\Delta a^{}_e$ is still allowed, namely, $\Delta a^{}_e \in [-0.34, 0.98]\times 10^{-12}$. The result in Eq.~(\ref{eq:DeltaaW}) is also applicable to the electron case, and one can estimate it by rescaling the value in Eq.~(\ref{eq:DeltaaWn}) by a factor of $m^2_e/m^2_\mu \approx 2.368\times 10^{-5}$. Therefore, the contribution of massive Majorana neutrinos to $\Delta a^{}_e$ is negligibly small. In our model, the interaction between the triplet Higgs and charged leptons can be described by
\begin{eqnarray}
{\cal L}^{}_{\rm Yuk} = - \frac{1}{2} y^{}_\Delta \left[ \sqrt{2} \left( \overline{e^{}_{\rm L}} \tau^{\rm C}_{\rm L} + \overline{\tau^{}_{\rm L}} e^{\rm C}_{\rm L} \right) H^{--} + \left(\overline{\nu^{}_{e{\rm L}}} \tau^{\rm C}_{\rm L} + \overline{\nu^{}_{\tau {\rm L}}} e^{\rm C}_{\rm L} + \overline{e^{}_{\rm L}} \nu^{\rm C}_{\tau{\rm L}} + \overline{\tau^{}_{\rm L}} \nu^{\rm C}_{e{\rm L}} \right) H^- \right] + {\rm h.c.} \; ,
\label{eq:YukawaDelta}
\end{eqnarray}
which induces the corrections to the anomalous magnetic moments of electron and tauon. The Feynman diagrams in the electron case have been shown in Fig.~\ref{fig:electrong}, and those in the tauon case can be obtained by exchanging $e^-$ and $\tau^-$. According to the formulas in Ref.~\cite{Moore:1984eg}, the contributions to $\Delta a^{}_e$ can be calculated as follows
\begin{eqnarray}
\Delta a^{--}_e &=& -\frac{y^2_\Delta m^2_e}{4\pi^2 M^2_\Delta} \int^1_0 {\rm d}x \left[\frac{x^2(1 - x)}{x + (m^2_\tau/M^2_\Delta) (1-x)} + \frac{x^2(1 - x)}{(1 - x) + (m^2_\tau/M^2_\Delta)x} \right] \approx - \frac{y^2_\Delta m^2_e}{12\pi^2 M^2_\Delta} \; , \nonumber \\
\Delta a^{-}_e &=& -\frac{y^2_\Delta m^2_e}{16\pi^2 M^2_\Delta} \int^1_0 {\rm d}x \frac{x^2(1 - x)}{x + (m^2_\tau/M^2_\Delta) (1-x)} \approx - \frac{y^2_\Delta m^2_e}{96\pi^2 M^2_\Delta} \; ,
\label{eq:aeDelta}
\end{eqnarray}
where the mass degeneracy of the triplet scalars is assumed and the terms proportional to small mass ratios $m^2_e/M^2_\Delta$ and $m^2_\tau/M^2_\Delta$ in the integrands have been omitted. The combination of these two contributions gives $\Delta a^{H}_e = \Delta a^{--}_e + \Delta a^-_e \approx - 3 y^2_\Delta m^2_e/(32\pi^2 M^2_\Delta)$. Since $y^{}_\Delta v^{}_\Delta \sim 0.1~{\rm eV}$ as indicated by the cosmological bound on the sum of three neutrino masses~\cite{Capozzi:2017ipn}, one gets $M^{}_\Delta v^{}_\Delta \sim 26.8~{\rm GeV}\cdot {\rm eV}$ in order to reach the lower bound of the $2\sigma$ range, i.e., $\Delta a^{}_e = -0.34 \times 10^{-12}$. At this point, we stress that the doubly-charged Higgs bosons $H^{\pm \pm}$ and their leptonic decays $H^{\pm \pm} \to e^\pm \tau^\pm$ can be observed at the CERN Large Hadron Collider. In particular, there exists only one leptonic decay channel $H^{\pm \pm} \to e^\pm \tau^\pm$, of which the total decay width is
\begin{eqnarray}
\Gamma(H^{\pm \pm} \to e^\pm \tau^\pm) = \frac{y^2_\Delta}{8\pi} M^{}_\Delta \; ,
\label{eq:Gamma}
\end{eqnarray}
where the dependence on the triplet Higgs mass $M^{}_\Delta$ is different from that in the case of $\Delta a^{H}_e$. Therefore, it is interesting to further investigate how to determine the model parameters $y^{}_\Delta$, $M^{}_\Delta$, and $v^{}_\Delta$ by combining neutrino oscillations, electron magnetic moment and the leptonic decays of doubly-charged Higgs bosons.
\begin{figure}[t!]
\begin{center}	
\hspace{0.5cm}	
\includegraphics[width=0.35\textwidth]{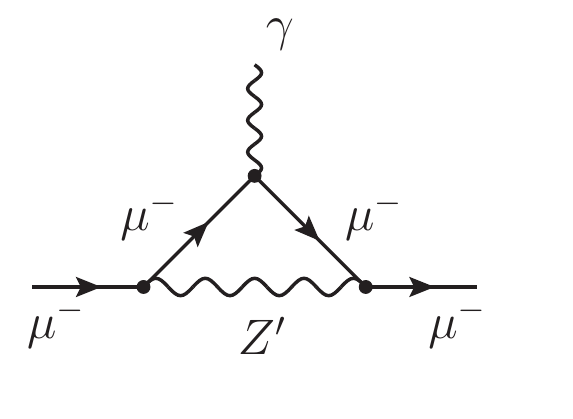}
\end{center}
\vspace{-1.2cm}
\caption{The Feynman diagram for the dominant contribution from $Z^\prime$ to the anomalous muon magnetic moment.}
\label{fig:Zprimeg} 
\end{figure}

Finally, we come to the anomalous muon magnetic moment, as shown in Fig.~\ref{fig:Zprimeg}, to which the dominant contribution is made by the new neutral gauge boson $Z^\prime$. Following the general formulas in Refs.~\cite{Leveille:1977rc, Moore:1984eg} and noticing the neutral-current interaction in Eq.~(\ref{eq:newNC}), we can calculate the correction from $Z^\prime$ to $\Delta a^{}_\mu$, namely,
\begin{eqnarray}
\Delta a^{Z^\prime}_\mu = \frac{g^2_{Z^\prime} m^2_\mu}{4\pi^2 M^2_{Z^\prime}} \int^1_0 {\rm d}x \frac{x^2(1-x)}{(m^2_\mu/M^2_{Z^\prime})x^2 + (1-x)} \approx \frac{g^2_{Z^\prime} m^2_\mu}{12\pi^2 M^2_{Z^\prime}} \; ,
\label{eq:amuZ}
\end{eqnarray}
where the term proportional to $m^2_\mu/M^2_{Z^\prime}$ has been omitted for $M^2_{Z^\prime} \gg m^2_\mu$ in the last step. As in our model, the gauge boson $Z^\prime$ acquires its mass $M^{}_{Z^\prime} = g^{}_{Z^\prime} v^{}_S/2$ after the spontaneous symmetry breaking. Therefore, if the anomalous muon magnetic moment $\Delta a^{}_\mu = 251\times 10^{-11}$ in Eq.~(\ref{eq:Deltaa}) is interpreted by the $Z^\prime$ contribution, then we can identify $\Delta a^{Z^\prime}_\mu$ with this number and estimate the vev of the singlet scalar from Eq.~(\ref{eq:amuZ}) as
\begin{eqnarray}
v^{}_S = \sqrt{\frac{m^2_\mu}{3\pi^2 \Delta a^{Z^\prime}_\mu}} \approx 385~{\rm GeV} \; .
\label{eq:vs}
\end{eqnarray}
In the ${\rm U}(1)_{L^{}_\mu - L^{}_\tau}$ gauge models, for $M^{}_{Z^\prime}$ around the GeV scale, the most stringent bound has been extracted from the measurement of neutrino trident events $\nu^{}_\mu + N \to \nu^{}_\mu + \mu^+ + \mu^- + N$~\cite{Altmannshofer:2014pba} using the data collected by the CCFR Collaboration~\cite{Mishra:1991bv}. In the mass range $M^{}_{Z^\prime} \in [0.4, 10^3]~{\rm GeV}$, the coupling $g^{}_{Z^\prime}$ favored by the muon $(g-2)$ has been excluded by the neutrino trident production~\cite{Altmannshofer:2014pba}. However, if $M^2_{Z^\prime} \ll m^2_\mu$ holds, then the result in Eq.~(\ref{eq:amuZ}) will be replaced by
\begin{eqnarray}
\Delta a^{Z^\prime}_\mu = \frac{g^2_{Z^\prime}}{8\pi^2} =  203\times 10^{-11} \cdot \left(\frac{g^{}_{Z^\prime}}{4\times 10^{-4}}\right)^2 \; .
\label{eq:amuZ2}
\end{eqnarray}
It is evident that the anomalous muon magnetic moment can be entirely explained by choosing $g^{}_{Z^\prime} \sim 4\times 10^{-4}$. For the intermediate range of $M^{}_{Z^\prime} \lesssim m^{}_\mu$, the loop integral in Eq.~(\ref{eq:amuZ}) varies from $0.1$ for $m^2_\mu/M^2_{Z^\prime} = 1$ to $1/2$ for $m^2_\mu/M^2_{Z^\prime} \gg 1$. One can easily find that $\Delta a^{Z^\prime}_\mu = 205\times 10^{-11}$ for $M^{}_{Z^\prime} = 105~{\rm MeV}$ and $g^{}_{Z^\prime} = 9\times 10^{-4}$. In this case, we have $v^{}_S = 2M^{}_{Z^\prime}/g^{}_{Z^\prime} \approx 233~{\rm GeV}$, which is related to other model parameters via $v^{}_\Delta = \lambda^{}_1 v^{}_S v^2_H/M^2_\Delta$. The constraints on $M^{}_{Z^\prime}$ at the MeV scale and the gauge coupling $g^{}_{Z^\prime}$ arise from terrestrial experiments~\cite{Kaneta:2016uyt, Araki:2017wyg, Zhang:2020fiu, Amaral:2021rzw}, astrophysical compact objects~\cite{Croon:2020lrf, KumarPoddar:2019ceq} and cosmology~\cite{Escudero:2019gzq}. However, the parameter space favoring the muon $(g-2)$ result still survives.

\section{Concluding Remarks}
\label{sec:conclu}

Motivated by the recent experimental measurement of the anomalous muon magnetic moment, which deviates from the theoretical prediction of the Standard Model at the $4.2\sigma$ level, we have proposed an extension of the type-(I+II) seesaw model by the gauged $L^{}_\mu - L^{}_\tau$ symmetry. By explicitly constructing a viable model with an additional singlet scalar, we demonstrate that the neutral gauge boson $Z^\prime$ associated with the ${\rm U}(1)^{}_{L^{}_\mu - L^{}_\tau}$ gauge symmetry can provide a possibility to explain the anomalous muon magnetic moment, while the lepton flavor structures are severely constrained as well. The main new results are summarized as below.

First, for the triplet Higgs $\Delta$ to develop a suitable vacuum expectation value, the introduction of a singlet scalar $S$ with an opposite charge under the ${\rm U}(1)^{}_{L^{}_\mu - L^{}_\tau}$ symmetry is necessary. At the same time, the mass $M^{}_{Z^\prime} = g^{}_{Z^\prime} v^{}_S/2$ of the neutral gauge boson $Z^\prime$ is essentially determined by the gauge coupling $g^{}_{Z^\prime}$ and the scale $v^{}_S$ of spontaneous gauge symmetry breaking. This is distinct from the phenomenological $Z^\prime$ models, where the gauge-boson mass $M^{}_{Z^\prime}$ and the gauge coupling $g^{}_{Z^\prime}$ are usually taken to be independent.

Second, thanks to the new flavor-dependent gauge symmetry, the lepton flavor structures have been severely restricted. The charged-lepton and Dirac neutrino Yukawa coupling matrices turn out to be flavor-diagonal. Meanwhile, the leptonic flavor mixing is mainly fixed by the structure of the right-handed neutrino mass matrix, receiving both contributions from the tree-level mass terms and the Yukawa interaction between right-handed neutrinos and the singlet scalar. Interestingly, the effective Majorana mass matrix of three ordinary neutrinos takes the form of the two-zero texture ${\bf B}^{}_3$, which is compatible with current neutrino oscillation data at the $3\sigma$ level. The predicted strong correlation between the neutrino mass ordering and the octant of $\theta^{}_{23}$ is readily testable in future neutrino oscillation experiments.

Third, we show that the anomalous magnetic moments of all three charged leptons receive contributions from massive Majorana neutrinos, which however are found to be negligibly small due to the experimental bounds on the unitarity violation of leptonic flavor mixing matrix. However, the anomalous electron magnetic moment also gets corrections from the Higgs triplet, and the correction with a negative sign is obtainable. The anomalous muon magnetic moment is dominantly explained by the radiative correction from the new neutral gauge boson $Z^\prime$. Taking into account the present experimental constraints, we find that the parameter space with $M^{}_{Z^\prime} \sim 100~{\rm MeV}$ and $g^{}_{Z^\prime} \sim 10^{-4}$ is still allowed.

We have to emphasize that the rich phenomenology of the proposed type-(I+II) model needs further dedicated investigations. The intrinsic correlation among the model parameters can be explored by carefully studying different physical processes, such as neutrino oscillations, lepton-flavor-violating decays of charged leptons, direct searches for the new particles at high-energy colliders, astrophysical and cosmological observations. If the reported discrepancy of anomalous muon magnetic moment is confirmed by the future precision data and refined calculations, it will be definitely intriguing to make a connection between the possible underlying new physics and the generation of neutrino masses and leptonic flavor mixing. We hope that the results in the present work will be instructive on this point. The systematic and self-consistent examination of all the phenomenological implications will be left for future works.

\section*{Acknowledgements}

The author thanks Yu-feng Li and Di Zhang for helpful discussions, and Prof. Zhi-zhong Xing for inspiring comments. This work was supported in part by the National Natural Science Foundation of China under Grant No.~11775232 and No.~11835013, and by the CAS Center for Excellence in Particle Physics. All the Feynman diagrams in this work were produced by using JaxoDraw~\cite{Binosi:2003yf}.

\end{document}